\documentclass[conference]{IEEEtran}
\IEEEoverridecommandlockouts

\usepackage{cite}
\usepackage{amsmath,amssymb,amsfonts}
\usepackage{graphicx}
\usepackage{textcomp}
\usepackage{xcolor}
\usepackage{amsmath,graphicx,hyperref}
\usepackage{amsmath,amssymb,amsfonts}
\usepackage{multirow}
\usepackage{graphicx} 
\usepackage{booktabs}
\usepackage{textcomp}
\usepackage{xcolor}
\usepackage{enumitem}
\usepackage{tikz}
\usetikzlibrary{shapes,arrows,positioning,fit,calc}
\usepackage{pgfplots}
\usepackage[most]{tcolorbox}
\pgfplotsset{compat=1.18}
\usetikzlibrary{shapes.geometric, arrows, positioning, calc}
\usetikzlibrary{arrows,decorations.pathmorphing,backgrounds,positioning,fit,petri}

\definecolor{echoreg}{HTML}{2cb1e1}
\definecolor{mymauve}{rgb}{0.58,0,0.82}
\definecolor{myorange}{rgb}{1.0,0.5,0.0}

\tikzset{
  cascaded/.style = {
    shadow scale=1,
    shadow xshift=2ex,
    shadow yshift=2ex,
    fill=white,
    draw=black,
    thick
  }
}
\usepackage{algorithm}
\usepackage{algpseudocode}
\usepackage{amsmath}
\usepackage{url} 
\usepackage{multirow}
\usepackage{graphicx}
\usepackage{tabularx}
\usepackage{booktabs}
\usepackage{makecell}
\usepackage{mdframed}

\def\BibTeX{{\rm B\kern-.05em{\sc i\kern-.025em b}\kern-.08em
    T\kern-.1667em\lower.7ex\hbox{E}\kern-.125emX}}
\begin{document}

\title{Weakly Supervised Detection and Temporal Localization of Whale Calls in Long-Duration Bioacoustic Data}


\author{\IEEEauthorblockN{Ragib Amin Nihal, Benjamin Yen, Runwu Shi, Takeshi Ashizawa, Kazuhiro Nakadai}
\IEEEauthorblockA{\textit{Systems and Control Engineering, School of Engineering, Institute of Science Tokyo, Japan} \\
}
}

\maketitle

\begin{abstract}
Passive acoustic monitoring (PAM) systems generate continuous recordings spanning months, yet automated bioacoustic analysis of whale calls requires two separate annotation efforts: binary presence labels for classification and precise temporal boundaries for localization. A binary label for a multi-minute recording can be assigned in seconds, but timestamping every call within it requires hours of expert effort. Providing both is infeasible at operational scale. We present DSMIL-LocNet, a weakly supervised multiple instance learning (MIL) framework that performs both classification and temporal localization using only recording-level presence/absence labels. Our dual-stream architecture integrates spectral and temporal features to process recordings of 2--30 minutes without the temporal compression that degrades existing CNN methods on long inputs. On the AcousticTrends BlueFinLibrary, DSMIL-LocNet achieves F1 scores of 0.88--0.91 on recordings of 300--1800s, where fully supervised CNN baselines degrade to 0.19--0.64. It also provides temporal localization that these baselines cannot produce without frame-level annotation. \\Code: \url{https://github.com/Ragib-Amin-Nihal/DSMIL-Loc}
\end{abstract}

\vspace*{-0.2cm}
\section{Introduction}
\label{sec:intro}

Passive Acoustic Monitoring (PAM) deploys hydrophones that record marine soundscapes continuously for months, generating terabytes of data per deployment in support of SDG~14 (\textit{Life Below Water})~\cite{mellinger2007overview}. Detecting whale vocalizations supports population monitoring and conservation. However, recordings arrive as multi-hour continuous streams rather than 
pre-segmented clips, making manual review impractical.

Early whale call detection methods focused on classification. Signal processing techniques included spectrogram matching~\cite{mellinger2000recognizing} and energy detection in specific frequency bands~\cite{mellinger2004comparison}. These methods worked for presence/absence decisions only in low-noise conditions. 

Deep learning models, primarily Convolutional Neural Networks (CNNs), 
have become the standard, improving classification accuracy 
for short ($\le$15 seconds) audio segments~\cite{allen2021convolutional,shiu2020deep,schall2024deep}; 
recent pre-trained foundation models~\cite{vanmerrienboer2026perch} extend 
this across taxa through transfer learning, but handle recording-level 
supervision via window-selection heuristics rather than learning temporal 
localization within long-duration streams. These methods inherit the same 
fundamental limitation: they solve only the \textbf{classification problem}. 
When researchers need \textbf{temporal localization}, existing methods 
require a separate annotation effort, with experts manually marking exact 
start and end times of every call. This creates two separate annotation 
tasks: labels for classification and precise timestamps for localization.
These two annotation tasks differ by orders of magnitude in cost: a binary 
label for a 30-minute recording takes seconds, while timestamping every 
call takes hours of expert effort~\cite{stowell2022computational}. For 
deployments producing thousands of hours, only recording-level labels are 
feasible at scale.
\begin{figure*}[t]
    \centering
    \includegraphics[width=0.75\linewidth]{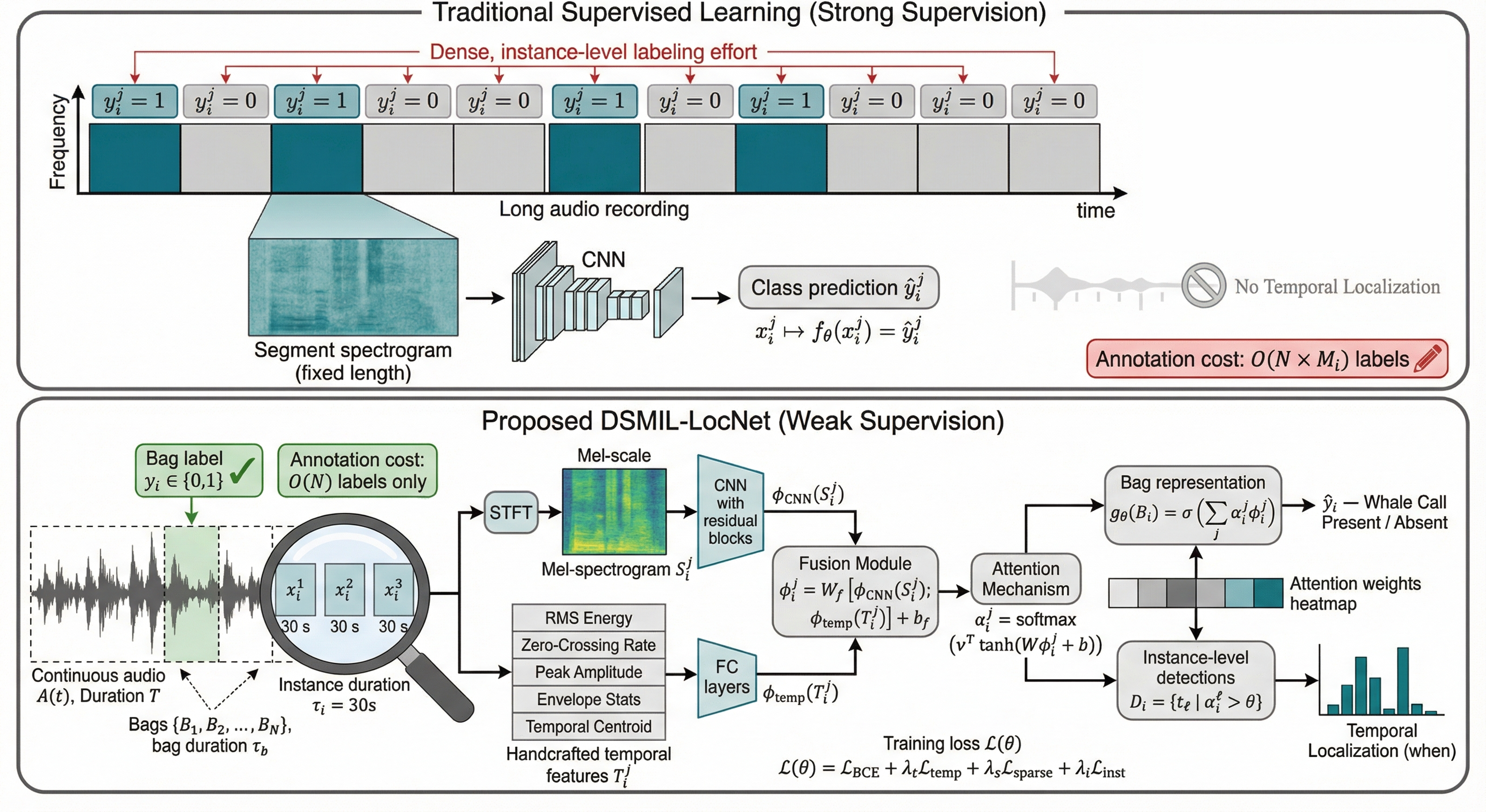}
    \vspace*{-3mm}
    \caption{
    \scriptsize{
    Comparison of supervision regimes. \textbf{Top:} The traditional approach requires instance-level annotations yet provides only classification with no temporal localization. \textbf{Bottom:} DSMIL-LocNet employs a Multiple Instance Learning framework that processes audio recordings as bags containing multiple instances. Spectrogram and temporal features are fed through a dual-stream architecture; the model achieves both classification and temporal localization of whale calls while requiring only \textbf{bag-level} annotations during training. 
    }}
    \label{fig:mil_architecture}
    \vspace*{-0.5cm}
\end{figure*}

Multiple Instance Learning (MIL) \cite{carbonneau2018multiple} offers an alternative framework for this problem.
Unlike supervised learning, MIL operates on ``bags" of instances, requiring labels only at the bag level (e.g., ``whale call present/absent" within a 1 hour recording). The instance-level labels (precise call timings) remain \textit{latent} or \textit{unknown} and not used during training, making it a weakly supervised learning approach. This characteristic of MIL is well-suited to the challenges of PAM data analysis, where we may know if whale calls are present within a longer recording segment (the bag), but lack precise annotations for each individual call event (the instances). 

MIL has been applied in audio tasks like acoustic scene classification \cite{choi2024instance} and sound event detection \cite{song2019acoustic}. However, these applications typically focus on classifying entire audio segments rather than localizing sound events within them. Moreover, most MIL approaches in audio do not explicitly tackle the challenges of long-duration recordings or the need for temporal localization. While MIL pooling, attention mechanisms, and multi-component losses are individually established, combining them for long-duration bioacoustic analysis introduces specific design challenges. Attention dilution at long durations requires explicit temporal smoothness and sparsity losses to keep the model focused on relevant segments. The dual-stream design addresses cases where spectral features alone lose temporal structure in long bags (see ablation, Table~\ref{tab:deployment_analysis}). Together, these components enable simultaneous classification and localization from weak supervision, a capability that none of them provides individually.

Unlike CNN attention mechanisms that operate on fixed-size inputs and require instance-level supervision, we hypothesize that MIL attention can handle variable-length sequences and learn to localize relevant instances from weak bag-level supervision alone. This motivates our central \textbf{research question}: 
\vspace*{-0.2cm}
\begin{mdframed}[
  leftline=true,
  rightline=false,
  topline=false,
  bottomline=false,
  linewidth=2pt,
]
\textit{Can a MIL architecture be designed that simultaneously achieves whale call classification and temporal localization using only binary bag-level labels, without additional temporal annotations?}
\end{mdframed}


To address this question, a Dual-Stream Multiple Instance Learning-Localization Network (DSMIL-LocNet, Figure \ref{fig:mil_architecture}) is introduced. It extracts both classification and localization information from weak supervision. Our method processes 2-30 minute recordings using only bag-level labels. The attention mechanisms for classification simultaneously learn temporal patterns, enabling localization without additional annotation. 
The \textbf{contributions} are:
\begin{enumerate}[leftmargin=*, nosep, label=(\arabic*)]
\item We formulate long-duration whale call analysis as a weakly supervised MIL problem that requires only a single binary label per multi-minute recording. A shared set of learned attention weights serves both bag-level classification and instance-level temporal localization, removing the dependency on frame-level annotations.
\item We propose a dual-stream architecture that combines spectral features from a CNN encoder with hand-crafted temporal features. Training uses a multi-component loss that accounts for attention dilution, class imbalance, and temporal coherence in long bags (ablated in Table~\ref{tab:deployment_analysis}).
\item We show that this approach maintains F1 scores of 0.88--0.91 on recordings of 300--1800s, where fully supervised CNNs degrade to 0.19--0.64, confirming its applicability to real PAM deployments.
\end{enumerate}
\section{Method}
\subsection{Problem Definition}
The key idea: we treat each long recording as a ``bag" of shorter segments (``instances"). During training, the model receives only a single label per bag (e.g., ``this 10-minute recording contains whale calls") but is never given \textit{which} segments contain calls. Through an attention mechanism, the model learns to assign high weights to segments likely containing calls and low weights to background. When aggregated, these weights produce the bag-level classification; individually, they provide temporal localization of calls within the recording, removing the need for frame-level temporal annotations.

Formally, let $\mathcal{A}(t)$ represent a continuous-time audio recording of duration $T$ minutes, segmented into $N$ non-overlapping bags $\{\mathcal{B}_1, \mathcal{B}_2, \dots, \mathcal{B}_N\}$ of duration $\tau_b$ minutes. Each bag $\mathcal{B}_i$ contains $M_i = \left\lfloor\frac{\tau_b}{\tau_i}\right\rfloor$ non-overlapping instances $\{\mathbf{x}_i^1, \mathbf{x}_i^2, \dots, \mathbf{x}_i^{M_i}\}$ of duration $\tau_i$. For each instance $\mathbf{x}_i^j$, timestamp $t_i^j$ records its start time. A binary bag-level label $y_i \in \{0, 1\}$ indicates whale vocalization presence, while instance-level labels $y_i^j \in \{0, 1\}$ remain \textit{unknown} during training. Following standard MIL assumptions:
\begin{align*}
y_i = 1 &\Leftrightarrow \exists j \in \{1, \dots, M_i\} : y_i^j = 1 \\
y_i = 0 &\Leftrightarrow \forall j \in \{1, \dots, M_i\} : y_i^j = 0
\vspace*{-0.2cm}
\end{align*}
Our goal is to learn a model parameterized by $\boldsymbol{\theta}$ that maximizes $\mathbb{P}(y_i | \mathcal{B}_i; \boldsymbol{\theta})$ subject to $\alpha_i^j \approx \mathbb{P}(y_i^j = 1 | \mathbf{x}_i^j; \boldsymbol{\theta})$, where $\alpha_i^j$ are attention weights corresponding to latent instance labels for temporal localization.
Each instance $\mathbf{x}_i^j$ is transformed into a $d$ dimension feature vector $\boldsymbol{\phi}_i^j \in \mathbb{R}^d$ via:
\[
\vspace*{-0.2cm}
\boldsymbol{\phi}_i^j = f(\mathbf{x}_i^j) = [\boldsymbol{\phi}_{{\text{CNN}}}(\mathbf{S}_i^j), \mathbf{T}_i^j].
\vspace*{-0.1cm}
\]
where
$\mathbf{S}_i^j$ is the time-frequency representation (Mel-spectrogram) computed from STFT, $\boldsymbol{\phi}_{{\text{CNN}}}$ extracts features from $\mathbf{S}_i^j$ via a CNN, $\mathbf{T}_i^j$ contains temporal features extracted from $\mathbf{x}_i^j$.
We define an instance-level prediction function, $h_{\boldsymbol{\theta}}(\mathbf{x}_i^j)$, with attention mechanism:
$h_{\boldsymbol{\theta}}(\mathbf{x}_i^j) = \sigma(\alpha_i^j \cdot \boldsymbol{\omega}^{Tr} \boldsymbol{\phi}(\mathbf{x}_i^j)),$
where $\sigma$ is the sigmoid function, $\boldsymbol{\omega}$ represents the learnable weights, $Tr$ is the transpose operator and $\boldsymbol{\phi}(\cdot)$ is the feature extraction function. The attention weight $\alpha_i^j$:
\[
\alpha_i^j = \frac{\exp(\mathbf{v}^{Tr} \tanh(\mathbf{W}\boldsymbol{\phi}(\mathbf{x}_i^j) + \mathbf{b}))}{\sum_{k=1}^{M_i} \exp(\mathbf{v}^{Tr} \tanh(\mathbf{W}\boldsymbol{\phi}(\mathbf{x}_i^k) + \mathbf{b}))} \cdot \mu_i^j.
\]
Here, $\mathbf{W}$, $\mathbf{v}$, and $\mathbf{b}$ are learnable, and $\mu_i^j$ is the binary mask for valid instances. The attention weights are normalized so that the sum of the weights in one bag equals to 1. 

The bag-level prediction, $g_{\boldsymbol{\theta}}(\mathcal{B}_i)$, is a weighted sum of instance embeddings,
$
g_{\boldsymbol{\theta}}(\mathcal{B}_i) =  \sum_{j=1}^{M_i} \alpha_i^j h_{\boldsymbol{\theta}}(\mathbf{x}_i^j).
$
For temporal localization, a detection is made when the attention weight exceeds threshold $\theta$:
$$\mathcal{D}_i = \{ t_i^j \mid \alpha_i^j > \theta, j \in \{1, \dots, M_i\} \}.$$
Given ground truth vocalization times $\mathcal{G}_i = \{g_i^1, g_i^2, \dots, g_i^{K_i}\}$, a detection $t_d \in \mathcal{D}_i$ matches a ground truth time $g_k \in \mathcal{G}_i$ if:
$\lvert t_d - g_k \rvert \leq \tau$, where $\tau$ is the temporal tolerance threshold and $\lvert \cdot \rvert$ denotes absolute value.
Let $\text{card}\{S\}$ denote the cardinality (number of elements) of set $S$. The localization performance is evaluated using:

\noindent\resizebox{\columnwidth}{!}{$
\vspace*{0.1cm}
\begin{aligned}
\text{Precision} &= \frac{\sum_i \text{card}\{ t_d \in \mathcal{D}_i \text{ s.t. } \exists g_k \in \mathcal{G}_i \text{ with } \lvert t_d - g_k \rvert \leq \tau \}}{\sum_i \text{card}\{\mathcal{D}_i\}}\\
\text{Recall} &= \frac{\sum_i \text{card}\{ g_k \in \mathcal{G}_i \text{ s.t. } \exists t_d \in \mathcal{D}_i \text{ with } \lvert t_d - g_k \rvert \leq \tau \}}{\sum_i \text{card}\{\mathcal{G}_i\}}
\end{aligned}
\vspace*{0.1cm}
$}
This defines the MIL problem for whale call localization, with feature extraction, attention mechanisms, and the relationship between instance-level and bag-level predictions.
\vspace*{-0.2cm}
\subsection{Bag and Instance Creation}
Algorithm \ref{alg:hierarchical} outlines the hierarchical segmentation of audio recordings into labeled bags of fixed-duration instances through resampling, bandpass filtering, and segmentation. Spectrogram and temporal features are extracted from each instance, and bag-level labels are assigned based on the presence of annotated whale calls.
\begin{algorithm}[t]
\caption{Hierarchical Audio Processing: }
\label{alg:hierarchical}
\small
\begin{algorithmic}[1]
\Require
    \State Audio recording $\mathcal{A}(t)$ with timestamps, Ground truth annotations $\mathcal{G}$, Bag duration $\tau_b$, Instance duration $\tau_i$, Target sampling rate $f_s$, Frequency range $[f_{min}, f_{max}]$

\Procedure{ProcessAcousticData}{$\mathcal{A}(t), \mathcal{G}$}
    \State $a \gets \textsc{Resample}(\mathcal{A}(t), f_s)$ 
    \State $a \gets \textsc{BandpassFilter}(a, f_{min}, f_{max})$ 
    \State $N \gets \left\lfloor\frac{\text{length}(a)}{f_s \times \tau_b}\right\rfloor$ \Comment{Number of bags}
    
    \For{$i \gets 1$ to $N$}
        \State $t_{start} \gets (i-1) \times \tau_b$
        \State $\mathcal{B}_i \gets \emptyset$, $y_i \gets 0$ \Comment{Initialize bag and label}
        \State $\text{bag\_data} \gets a[t_{start} \times f_s : (t_{start} + \tau_b) \times f_s]$
        \State $M_i \gets \left\lfloor\frac{\tau_b}{\tau_i}\right\rfloor$ \Comment{Instances per bag}
        
        \For{$j \gets 1$ to $M_i$}
            \State $t_i^j \gets t_{start} + (j-1) \times \tau_i$ \Comment{timestamp}
            \State $\mathbf{x}_i^j \gets \text{bag\_data}[(j-1) \times \tau_i \times f_s : j \times \tau_i \times f_s]$
            
            \State $S_i^j \gets \textsc{STFT}(\mathbf{x}_i^j)$, 
            \State $\boldsymbol{\phi}_{\text{CNN}} \gets \textsc{MelTransform}(S_i^j)$
            \State $\mathbf{T}_i^j \gets \textsc{Temporal}(\mathbf{x}_i^j)$, $\boldsymbol{\phi}_i^j \gets [\boldsymbol{\phi}_{\text{CNN}}, \mathbf{T}_i^j]$
            
            \If{\textsc{Overlap}($t_i^j, t_i^j + \tau_i, \mathcal{G}$)} $y_i \gets 1$ \EndIf
            \State $\mathcal{B}_i \gets \mathcal{B}_i \cup \{(\mathbf{x}_i^j, \boldsymbol{\phi}_i^j, t_i^j)\}$
        \EndFor
        
        \State $\mathcal{B} \gets \mathcal{B} \cup \{(\mathcal{B}_i, y_i)\}$
    \EndFor
    \State \Return $\mathcal{B}$
\EndProcedure
\end{algorithmic}
\end{algorithm}

\vspace*{-0.2cm}
\subsection{Network Architecture}
DSMIL-LocNet extracts both frequency and time-domain characteristics of whale vocalizations. For each instance $\mathbf{x}_i^j$, we compute a time-frequency representation,
$\mathbf{S}_i^j = \text{MelScale}(|\text{STFT}(\mathbf{x}_i^j)|^2).$
This transformation retains whale call frequency modulation while reducing dimensionality. Temporal features $\mathbf{T}_i^j$ are extracted as:
\vspace*{-0.2cm}
\begin{equation*}
\resizebox{\columnwidth}{!}{%
$\mathbf{T}_i^j =
\begin{bmatrix}
\text{RMS}(\mathbf{x}_i^j), & \text{ZCR}(\mathbf{x}_i^j), & \text{PeakAmp}(\mathbf{x}_i^j), & \text{EnvStats}(\mathbf{x}_i^j), & \text{TempCentroid}(\mathbf{x}_i^j)
\end{bmatrix}$%
}
\vspace*{-0.2cm}
\end{equation*}
where RMS represents root mean square energy, ZCR is the zero-crossing rate, and TempCentroid calculates the temporal center of mass of the signal, along with peak amplitude and envelope statistics (mean, standard deviation). These features capture the temporal structures that distinguish whale calls from background noise.
Our dual-stream architecture processes spectral and temporal features separately. The spectrogram stream $\boldsymbol{\phi}_{{\text{CNN}}}$ employs residual blocks:
\vspace*{-0.2cm}
\begin{equation*}
\begin{aligned}
\mathbf{h}_{l+1} &= \mathbf{h}_l + F(\mathbf{h}_l), \\
\text{where } \quad
F(\mathbf{h}_l) &= \mathbf{W}_2 \sigma\!\left(
\text{BN}\!\left(
\mathbf{W}_1 \sigma\!\left(
\text{BN}(\mathbf{h}_l)
\right)
\right)
\right).
\end{aligned}
\vspace*{-0.2cm}
\end{equation*}

where $\mathbf{h}_l$ represents the feature maps at layer $l$, and $F(\cdot)$ is the residual function, $\mathbf{W}_1$ and $\mathbf{W}_2$ are convolutional weight matrices, BN denotes batch normalization, and $\sigma$ is the ReLU activation function. 
The residual connections maintain gradient flow through deep layers. The temporal stream features:
\vspace*{-0.2cm}
\begin{equation*}
\boldsymbol{\phi}_{\text{temp}}(\mathbf{T}_i^j) = \sigma(\mathbf{W}_n\text{BN}(\sigma(\mathbf{W}_{n-1}...\sigma(\mathbf{W}_1\mathbf{T}_i^j)))).
\vspace*{-0.2cm}
\end{equation*}
Our feature fusion module learns to dynamically weight and combine information from both streams:
\vspace*{-0.2cm}
\begin{equation*}
\boldsymbol{\phi}_i^j = \mathbf{W}_f[\boldsymbol{\phi}_{{\text{CNN}}}(\mathbf{S}_i^j); \boldsymbol{\phi}_{{\text{temp}}}(\mathbf{T}_i^j)] + \mathbf{b}_f.
\vspace*{-0.2cm}
\end{equation*}
This fusion allows the model to adapt its feature emphasis based on the signal characteristics, particularly important when dealing with varying noise conditions and call types.
Instance-level attention weight  $\alpha_i^j = \text{softmax}(\mathbf{v}^{Tr}\tanh(\mathbf{W}\boldsymbol{\phi}_i^j + \mathbf{b})),$
where the softmax operation normalizes attention weights across all instances within bag $i$. During inference, the model produces both bag-level predictions:
$g_{\boldsymbol{\theta}}(\mathcal{B}_i) = \sigma\left(\sum_j \alpha_i^j \boldsymbol{\phi}_i^j\right)$ and instance-level temporal localization through $\alpha^j_i$, providing a framework for both tasks.
\vspace*{-0.2cm}
\subsection{Loss Function}
\noindent Our loss function $\mathcal{L}(\boldsymbol{\theta})$ comprises four components addressing class imbalance, temporal coherence, sparsity, and instance consistency.
The loss uses focal binary cross-entropy:
\begin{equation}
\vspace*{-0.2cm}
    \mathcal{L}_{\text{BCE}}(y_i,\hat{y}_i) = -y_i(1-\hat{y}_i)^\gamma \log(\hat{y}_i) - (1-y_i)\hat{y}_i^\gamma \log(1-\hat{y}_i).
    \label{eq:bce}
\end{equation}
Where $\gamma$ is the focal loss parameter that down-weights easy examples to address class imbalance in whale call detection with rare positives. Label smoothing with factor $\rho$ adjusts targets to $y_i' = y_i(1-\rho) + \rho/2$.

We add a temporal smoothness constraint for coherent attention across instances, reflecting whale call continuity:
\vspace*{-0.2cm}
$$\mathcal{L}_{\text{temp}} = \frac{1}{M_i} \sum_i \sum_j (\alpha_i^j - \alpha_i^{j-1})^2.\vspace*{-0.2cm}$$
As whale calls occupy a small fraction, to ensure selective attention and consistent feature representations, we include sparsity and instance consistency terms:
$$\vspace*{-0.1cm}\mathcal{L}_{{\text{sparsity}}} = \frac{1}{M_i} \sum_i \sum_j |\alpha_i^j|,
\mathcal{L}_{{\text{inst}}} = \frac{1}{M_i^2} \sum_i \sum_{j,k} \|\boldsymbol{\phi}_i^j - \boldsymbol{\phi}_i^k\|_2^2.\vspace*{-0.2cm}$$
The total loss is a weighted combination of these components:
\vspace*{-0.2cm}
\begin{equation}
\mathcal{L}(\boldsymbol{\theta}) = \mathcal{L}_{{\text{BCE}}} + \lambda_t \mathcal{L}_{{\text{temp}}} + \lambda_s \mathcal{L}_{{\text{sparsity}}} + \lambda_i \mathcal{L}_{{\text{inst}}}.
\label{eq:total}
\vspace*{-0.2cm}
\end{equation}
where $\lambda_t$, $\lambda_s$, and $\lambda_i$ are hyperparameters controlling the contribution of each loss component.
\section{Experiments, Results \& Discussion}
\begin{figure}[t]
    \centering
    \includegraphics[width=0.95\linewidth]{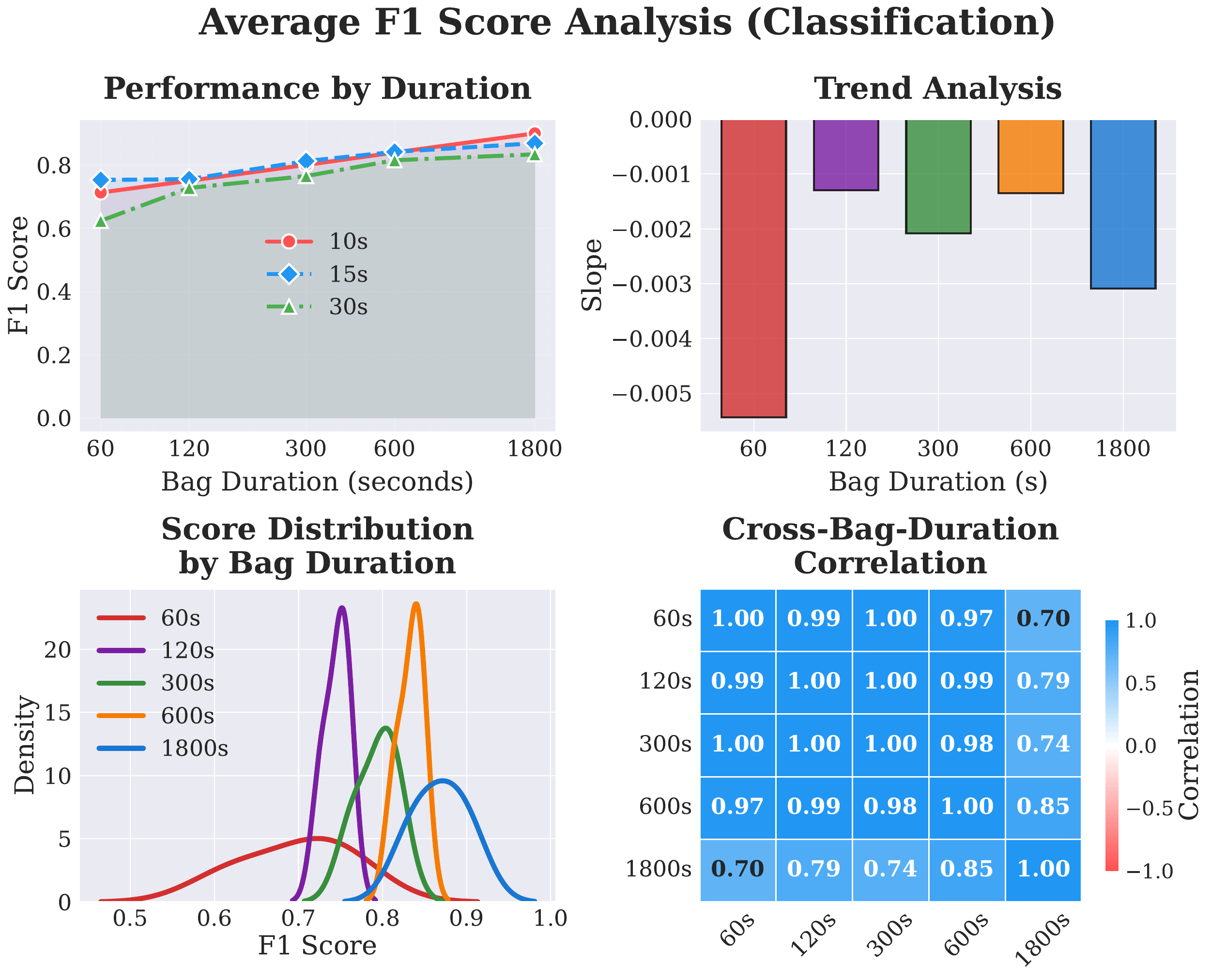}
    \caption{Classification F1 improves with bag duration, with 
the gains between 60\,s and 300\,s. The 15\,s instance 
length performs best across all durations.}
    \label{fig:f1_scores_analysis}
    \vspace*{-0.6cm}
\end{figure}
\subsection{Experimental Setup}
We evaluated our method using the AcousticTrends BlueFinLibrary \cite{miller2021open}, which contains 1880.25 hours of annotated Antarctic whale recordings from seven sites across 11 site-year combinations. The site-years have background noise comprising 73-99.9\% of recordings. Blocked cross-validation where each site-year serves as an independent test set. Implementation uses PyTorch with AdamW optimizer (lr=1e-4), focal loss ($\gamma$=2.0), early stopping (patience=15), and batch size 32 on NVIDIA A100 GPUs. F1-score for classification and precision for temporal localization is used. We emphasize precision for localization because, in operational PAM workflows, flagged segments can be reviewed by human analysts. High precision ensures that flagged time windows predominantly contain true calls, minimizing wasted review effort. Missed calls (lower recall) are less critical in this context, as recordings can be re-analyzed; false alarms consume limited expert time.
\subsection{Framework Comparison}
We first examine the core architectural differences that define each method's capabilities, summarized in Table~\ref{tab:method_comparison}.

\textbf{CNN-based approaches} (ANIMAL-SPOT~\cite{bergler2022animal}, DeepWhaleNet~\cite{rasmussen2024deepwhalenet}, Koogu~\cite{madhusudhana2022koogu}) process spectrograms as fixed-size images, requiring temporal compression that eliminates resolution needed for localization. Recent work has also explored foundation-model approaches 
to audio segmentation~\cite{audiosam2024}. These methods operate on pre-segmented, short-duration inputs 
with full supervision and do not address joint 
classification--localization under weak supervision.
\textbf{Template-based methods} (WT-HMM~\cite{babalola2024wavelet}) offer limited localization through pattern matching but lack adaptability to diverse call types and noise conditions encountered in real environments.
\textbf{DSMIL-LocNet} overcomes these limitations through MIL, processing variable-length recordings without compression while providing adaptive temporal localization.
\subsection{Cross-Method Performance Analysis}
Table~\ref{tab:method_comparison} compares detection performance across durations from 60\,s to 1800\,s. The central finding is that DSMIL-LocNet sustains high performance (F1: 0.88--0.91) at 300--1800\,s using only bag-level labels, while all fully supervised baselines degrade to F1 0.19--0.64 at these durations due to information loss from temporal compression. Furthermore, DSMIL-LocNet provides temporal localization across all durations---a capability that none of the supervised baselines offer without additional annotation.

At short durations (60--120\,s), fully supervised CNNs outperform our weakly supervised method, which is expected: these models have access to dense instance-level labels and are optimized for classifying individual, pre-segmented calls. The operationally relevant comparison is at longer durations that reflect real PAM data, where recordings arrive as continuous multi-minute streams and per-instance annotation is infeasible.

\begin{figure}[t]
    \centering
    \includegraphics[width=0.95\linewidth]{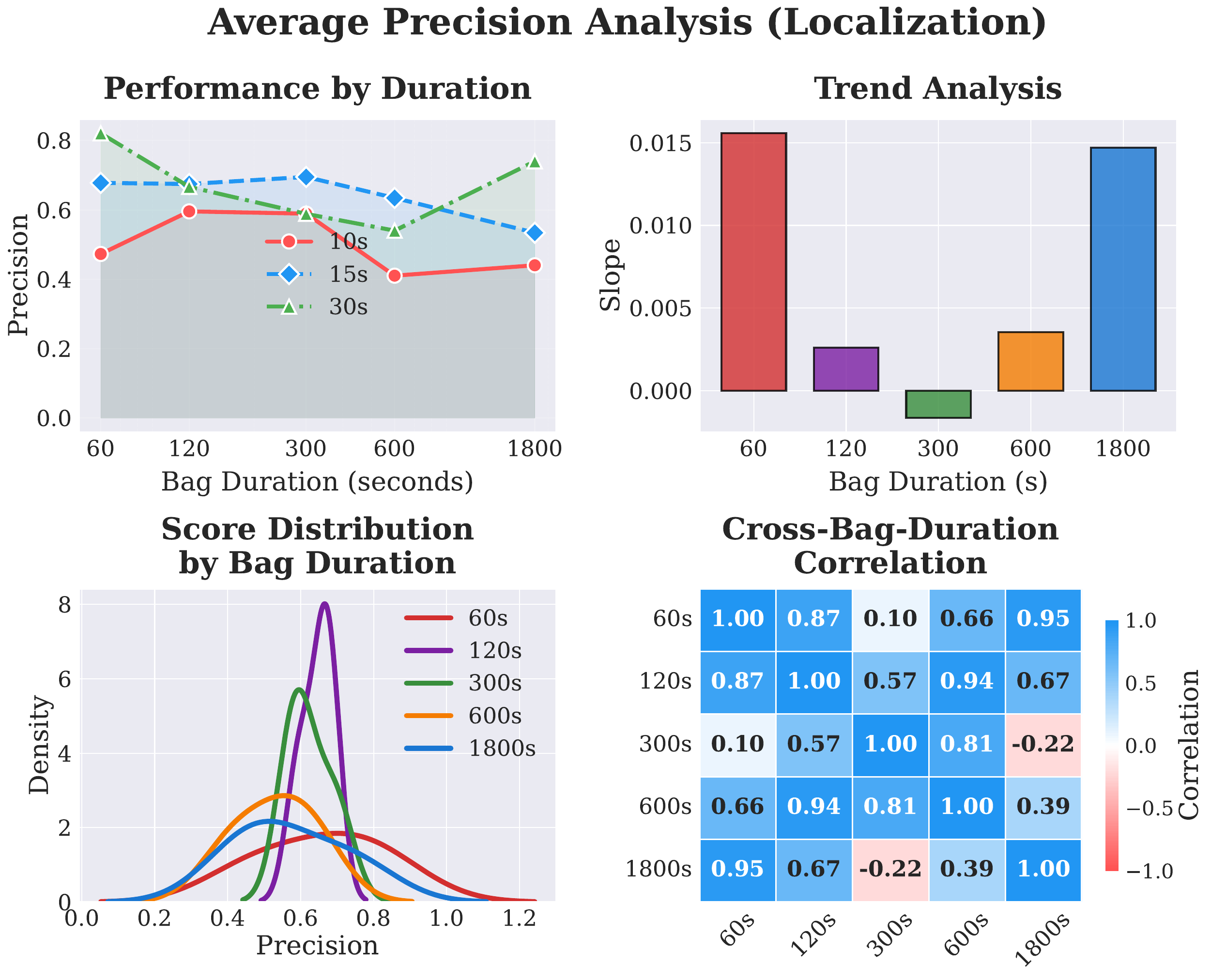}
    \caption{Localization precision peaks at short bags (60--120\,s) 
and degrades at longer durations due to attention dilution 
across more instances.}
    \label{fig:precision_scores_analysis}
\vspace*{-0.3cm}
\end{figure}
\begin{table*}[t]
\caption{Method Comparison: Architecture and Performance.\\DSMIL-LocNet uses only weak bag-level labels; all baselines require strong supervision with instance-level annotations}
\label{tab:method_comparison}
\vspace*{-0.2cm}
\centering
\resizebox{\textwidth}{!}{%
\begin{tabular}{l c c c c c c c c c}
\toprule
\multirow{2}{*}{Method} & \multirow{2}{*}{\makecell{Input\\Processing}} & \multirow{2}{*}{\makecell{Localization\\Method}} & \multirow{2}{*}{Supervision} & \multicolumn{5}{c}{Classification F1 Score} & \multirow{2}{*}{\makecell{Temporal\\Localization}} \\
\cmidrule(lr){5-9}
& & & & 60s & 120s & 300s & 600s & 1800s & \\
\midrule
\textbf{DSMIL-LocNet} & Hierarchical & Attention weights & \textbf{Weak} & 0.75 & 0.79 & \textbf{0.88} & \textbf{0.89} & \textbf{0.91} & \textbf{Yes} \\
\midrule
ANIMAL-SPOT~\cite{bergler2022animal} & Fixed 2D CNN & Not available & Strong & \textbf{0.83} & \textbf{0.82} & 0.64 & 0.35 & 0.22 & No \\
DeepWhaleNet~\cite{rasmussen2024deepwhalenet} & Fixed ResNet & Not available & Strong & 0.78 & 0.72 & 0.58 & 0.31 & 0.19 & No \\
WT-HMM~\cite{babalola2024wavelet} & Template matching & Template correlation & Strong & 0.65 & 0.58 & 0.42 & 0.22 & 0.16 & Limited\\
Koogu (DenseNet)~\cite{madhusudhana2022koogu} & Fixed DenseNet & Not available & Strong & 0.59 & 0.52 & 0.38 & 0.24 & 0.18 & No \\
\bottomrule
\end{tabular}%
}
\vspace*{-0.4cm}
\end{table*}
\vspace*{-0.2cm}
\subsection{Optimization to Long-Duration Processing}
We analyzed how bag and instance duration parameters affect classification and localization performance.
As shown in Figure~\ref{fig:f1_scores_analysis}, classification F1 improves with longer bag durations, which explains the model's advantage on segments over 300s: longer bags better handle class imbalance by increasing the likelihood that a bag contains at least one whale call.

Figure~\ref{fig:precision_scores_analysis} shows an inverse pattern for localization. 
Because attention weights are softmax-normalized across all instances in a bag, so as bag length grows, the number of competing instances increases (from 4 at 60s to 120 at 1800s, with 15s instances), diluting the attention assigned to any individual call-bearing instance. This is a property of softmax-normalized attention over variable-length sequences; any attention-based MIL method operating over variable bag lengths will exhibit the same behavior.

Two points are worth noting regarding this tradeoff. First, even at 600s where localization precision drops to 0.5--0.6, fully supervised baselines provide no localization at any duration, since they lack the mechanism to produce temporal predictions. Reduced-precision localization from weak supervision is still more informative than no localization from strong supervision. Second, the tradeoff suggests a two-stage deployment workflow: use long bags (300--1800s) for high-accuracy classification to identify recordings containing calls, then re-segment only the positive recordings into shorter bags (60--120s) for precise localization. Both stages require only bag-level labels and no retraining.
\vspace*{-0.1cm}
\subsection{Ablation Analysis}
Table \ref{tab:deployment_analysis} demonstrates each architectural component's contribution. 
The ablation study verifies that the full dual-stream architecture with multi-component loss is necessary for optimal performance. This approach successfully bridges the gap between short-segment model optimization and the long-duration requirements of marine monitoring.
\vspace*{-0.1cm}
\begin{table}[t]
\caption{Ablation Study Results}
\label{tab:deployment_analysis}
\vspace*{-0.2cm}
\centering
\resizebox{\columnwidth}{!}{%
\begin{tabular}{l c c c c}
\toprule
Configuration & Loss Function & F1-Class. & Temp. Loc. Precision & \\
\midrule
Temporal-only MIL & $\mathcal{L}_{\text{BCE}}$ & 0.1235 & 0.4173 & \\
Spectrogram-only MIL & $\mathcal{L}_{\text{BCE}}$ & 0.7391 & 0.5371 & \\
Basic Dual-Stream MIL & $\mathcal{L}_{\text{BCE}}$ & 0.7880 & 0.6352 & \\
\textbf{DSMIL-LocNet (Full)} & $\mathcal{L}(\boldsymbol{\theta})$ & \textbf{0.8133} & \textbf{0.6952} & \\
\bottomrule
\end{tabular}%
}
\vspace{-0.4cm}
\end{table}
\section{Conclusion}
\vspace*{-0.05cm} 
This work shows that in long-duration acoustic monitoring, weak supervision can provide capabilities that strong supervision with fixed-input architectures does not support. DSMIL-LocNet performs both classification and temporal localization from a single binary label per recording, maintaining F1 scores of 0.88--0.91 at 300--1800s where fully supervised baselines fall to 0.19--0.64. In practice, this means an analyst labeling at the bag level can generate training data far more efficiently, while the model produces both detection and localization outputs that previously required separate annotation efforts. The underlying principle is not specific to marine bioacoustics. Monitoring domains characterized by continuous data streams, sparse target events, and high annotation costs (e.g., seismology, medical acoustics, terrestrial biodiversity surveys) face similar supervision constraints. Methods that convert inexpensive labels into structured temporal predictions are likely to be useful in these settings as well, particularly as large-scale sensor networks continue to grow.
\vspace*{-0.2cm}
\section*{Acknowledgment}
This work was supported in part by JSPS KAKENHI Grant Number JP20H00475 
and used the TSUBAME4.0 supercomputer at Institute of Science Tokyo.
\bibliographystyle{IEEEbib}
\bibliography{ref}

\end{document}